\documentclass[twocolumn]{aastex631}

\usepackage{color}
\usepackage{url}
\usepackage{hyperref}
\usepackage{xspace}
\usepackage{soul}

\newcommand{\psrtar}{MAXI\,J1816$-$195}

\newcommand{\nustar}{{NuSTAR}\xspace}
\newcommand{\nicer}{{NICER}\xspace}  
\newcommand{\swift}{{Swift}\xspace}
\newcommand{\rxte}{{RXTE}\xspace}
\newcommand{\hxmt}{{Insight-HXMT}\xspace}

\def\chiq{$\chi^2$}

\def\be{\begin{equation}} 
\def\ee{\end{equation}}

\begin{document}

\title{Broadband X-ray timing and spectral characteristics of the accretion-powered millisecond X-ray pulsar \psrtar}

\correspondingauthor{Zhaosheng Li, Mingyu Ge}
\email{lizhaosheng@xtu.edu.cn, gemy@ihep.ac.cn}

\author[0000-0003-2310-8105]{Zhaosheng Li}
\affiliation{Key Laboratory of Stars and Interstellar Medium, Xiangtan University, Xiangtan 411105, Hunan, China}

\author[0000-0002-7889-6586]{Lucien Kuiper}
\affiliation{SRON-Netherlands Institute for Space Research, Niels Bohrweg 4, 2333 CA, Leiden, The Netherlands}

\author[0000-0002-3776-4536]{Mingyu Ge}
\affiliation{Key Laboratory of Particle Astrophysics, Institute of High Energy Physics, Chinese Academy of Sciences, 19B Yuquan Road, Beijing 100049, China}

\author[0000-0003-3095-6065]{Maurizio Falanga}
\affiliation{International Space Science Institute (ISSI), Hallerstrasse 6, 3012 Bern, Switzerland}
\affiliation{Physikalisches Institut, University of Bern, Sidlerstrasse 5, 3012 Bern, Switzerland}

\author[0000-0002-0983-0049]{Juri Poutanen}
\affiliation{Department of Physics and Astronomy, University of Turku, FI-20014, Finland}

\author[0000-0001-9599-7285]{Long Ji}
\affiliation{School of Physics and Astronomy, Sun Yat-sen University, Zhuhai, 519082, People’s Republic of China}

\author{Yuanyue Pan}
\affiliation{Key Laboratory of Stars and Interstellar Medium, Xiangtan University, Xiangtan 411105, Hunan, P.R. China}

\author{Yue Huang}
\affiliation{Key Laboratory of Particle Astrophysics, Institute of High Energy Physics, Chinese Academy of Sciences, 19B Yuquan Road, Beijing 100049, China}

\author[0000-0002-9042-3044]{Renxin Xu}
\affiliation{Department of Astronomy, School of Physics, Peking University, Beijing 100871, China}
\affiliation{Kavli Institute for Astronomy and Astrophysics, Peking University, Beijing 100871, China}

\author[0000-0003-0274-3396]{Liming Song}
\affiliation{Key Laboratory of Particle Astrophysics, Institute of High Energy Physics, Chinese Academy of Sciences, 19B Yuquan Road, Beijing 100049, China}

\author{Jinlu Qu}
\affiliation{Key Laboratory of Particle Astrophysics, Institute of High Energy Physics, Chinese Academy of Sciences, 19B Yuquan Road, Beijing 100049, China}

\author{Shu Zhang}
\affiliation{Key Laboratory of Particle Astrophysics, Institute of High Energy Physics, Chinese Academy of Sciences, 19B Yuquan Road, Beijing 100049, China}

\author{Fangjun Lu}
\affiliation{Key Laboratory of Particle Astrophysics, Institute of High Energy Physics, Chinese Academy of Sciences, 19B Yuquan Road, Beijing 100049, China}

\author[0000-0001-5586-1017]{Shuang-Nan Zhang}
\affiliation{Key Laboratory of Particle Astrophysics, Institute of High Energy Physics, Chinese Academy of Sciences, 19B Yuquan Road, Beijing 100049, China}



\begin{abstract}
%We studied the broadband timing and spectral behaviors of the newly confirmed accreting millisecond X-ray pulsar \psrtar\ in the 2022 outburst. We used the data from \nicer, \nustar, and \hxmt\ ME/HE, which cover the energy range between 0.8--210 keV. We detected the X-ray pulsation up to $\sim95$ keV. 
%The pulse profiles were quite stable over the whole outburst and can be well described by a truncated Fourier series using two harmonics, the fundamental and the first overtone. Both components were well aligned in the range 0.8--64 keV.   The coherent timing analysis of \nicer\ and \hxmt\ ME/HE observations reveals a complex behavior between MJD 59737.0-59741.9 and  significant spin-up, $\dot{\nu}=(9.0\pm2.1)\times10^{-14}~{\rm Hz~s^{-1}}$, between MJD 59741.9--59760.6.
%The joint and time-averaged \nicer\ and \hxmt\ spectra in the energy range 1--150 keV are well fitted by the absorbed Comptonization model {\sc compps} plus disk blackbody with additional two Gaussian components. Based on the bolometric flux during the spin-up epoch, we determine the magnetic field strength of $(0.23-1.11)\times10^8$ G in \psrtar.
%The \nustar\ spectra, however, showed the reflection from accretion disk.
We studied the broadband X-ray timing and spectral behaviors of the newly confirmed accreting millisecond X-ray pulsar \psrtar\ during its 2022 outburst. 
We used the data from \hxmt\ ME/HE, \nicer\ and \nustar\ which cover the energy range between 0.8--210 keV. 
A coherent timing analysis of solely \hxmt\ HE data across the full outburst revealed a complex behavior of the timing residuals, also prominently visible in independent \hxmt\ ME and \nicer\ data, particularly at rising part of the outburst and at the very end in \nicer\ data. Therefore, we broke down the full outburst
into a (noisy) rising part, covering only about five days from MJD 59737.0 to 59741.9, and a decaying part lasting for 19 days across MJD 59741.9--59760.6. 
%%, not covering the end phase of the outburst for which \nicer\ sees deviant behaviour because the lack of \hxmt\ measurements. 
Fitting for the decaying part a timing model including a frequency $\nu$ and frequency time derivative $\dot{\nu}$ component yielded a value of $(+9.0\pm2.1)\times10^{-14}~{\rm Hz~s^{-1}}$ for $\dot{\nu}$, which could be interpreted as a spin-up under our model assumptions.
We detected the X-ray pulsations up to $\sim$95~keV in a combination of
\hxmt\ HE observations. The pulse profiles were quite stable over the whole outburst and could be well described by a truncated Fourier series using two 
harmonics, the fundamental and the first overtone. Both components kept alignment in the range 0.8--64~keV. The joint and time-averaged \nicer\ and \hxmt\ 
spectra in the energy range 1--150~keV were well fitted by the absorbed Comptonization model {\tt compps} plus disk blackbody with two additional Gaussian 
components. Using the bolometric flux and spin-up values both evaluated during the decay phase, we determined a magnetic field strength of $(0.2-2)\times10^8$ G for \psrtar.
 
\end{abstract}



\section{Introduction} \label{sec:intro}

% After the discovery of radio millisecond pulsar, the recycling scenario has been proposed to explain 

% Recycling scenario of radio millisecond X-ray pulsar...

% Burst oscillation...

% Accretion millisecond X-ray pulsar...

% Spin-up during outburst...

% Switching between accretion-power and rotation-power...

% Magnetic field...

As a  subclass of neutron star low-mass X-ray binary (NS LMXB), accreting millisecond X-ray pulsars (AMXPs) are usually confirmed by its coherent pulsation during outburst, with spin periods of a few milliseconds and orbital periods between 40~min and 5~h based on the current sample \citep[see][for reviews]{Campana2018,DiSalvo2020,papitto20,Patruno2021}. The recycling scenario suggested that AMXPs are progenitors of (radio) millisecond pulsars \citep{Alpar82,r82}. 
This picture has been verified by the discoveries of AMXPs starting with SAX J1808.4--3658 by \rxte\ \citep{Wijnands98}, discovery of the pulsar spin-up induced by accretion \citep{falanga05}, and by  swinging between accretion- and rotation-powered states in IGR J18245--2452 \citep{papitto13c}.

The new X-ray transient \psrtar\ was discovered by MAXI/Gas Slit Camera on June 7 during its 2022 outburst \citep{ATel15418}. Using the \swift\ localization, \nicer\ detected pulsations and thermonuclear type I X-ray bursts from \psrtar, confirming that the source is an AMXP with a spin frequency of 528 Hz, an orbital period of 4.83 hr and a projected semi-major axis of 0.26 lt-s \citep{ATel15421,ATel15425,ATel15431, Bult22}. The mass of the companion star was determined to be in the range 0.10--0.55~$M_\odot$ \citep{Bult22}.    Meanwhile, \hxmt\ found X-ray pulsations from \psrtar\ in the hard X-ray/soft $\gamma$-ray band \citep{ATel15471}. \citet{Bult22} proposed the flux-bias model, which accounts for the effects of the accretion torque and/or the wandering of the hot spot on the NS surface, to explain the structures left in the timing residuals across the outburst.   \citet{Chen2022} reported the detection of 73 type I X-ray bursts by \hxmt\ from \psrtar, and obtained an upper limit of the distance as 6.3 kpc. Wang et al. (2023, in preparation) found the type I X-ray bursts rate decreasing for the bolometric flux higher than $\sim1.04\times10^{-8} \rm{erg~cm^{-2}~s^{-1}}$ and estimated the distance to the source at $\sim3.4$ kpc \citep{Cavecchi20}.

% a notable hard X-ray shortage during bursts, indicating cooling of the corona. 
%Moreover, the frequently and quasi-periodic type I X-ray bursts have the recurrence time between $1.15-2$ hours (Wang et al. 2022, ApJ submitted).   

% \psrtar\ is also a peculiar X-ray burster.

Since the onset of the 2022 outburst a search for radio, optical and near-infrared emission from \psrtar\ had been carried out. The radio emission of \psrtar\ showed significant evolution, at 1.28 GHz with a flux density of  $4.2\pm0.4~{\rm mJy}$ on 2022 June 9 and a $3\sigma$ upper limit of $70{~\rm \mu Jy}$ on 2022 June 11 by MeerKAT, at 5.5 GHz with a flux density of $135\pm35~{\rm \mu Jy}$ on 2022 June 12 by the Australia Telescope Compact Array (ATCA), and at 6 GHz with a flux density of  $457\pm20~{\rm \mu Jy}$ on 2022 June 18 by VLA \citep{ATel15481, ATel15484}. However, radio pulsations were not detected. The likely optical and near-infrared counterpart of \psrtar\ had been identified  \citep{ATel15479,ATel15501}. When \psrtar\ evolved towards the quiescent state about one month after the start of the outburst, X-ray emission from a dust scattered halo around the source position was detected by \swift\ \citep{ATel15506}.

In this work, we present the broad band timing and spectral analysis of \psrtar. We introduce the instruments, \hxmt, \nicer\ and \nustar, and the performed observations in Sect.~\ref{sec:obs}. 
The timing and the broad-band spectral results are reported in Sect.~\ref{sec:timing} and Sect.~\ref{sec:spec}, respectively, and our findings are discussed in Sect.~\ref{sec:discussion}.

\section{Observations and data reduction}\label{sec:obs}

We analyzed the data from \psrtar\ collected by \hxmt, \nicer, and \nustar. 

\subsection{\hxmt}\label{sec:hxmt}

\hxmt\ \citep[Insight Hard X-ray Modulation Telescope,][]{hxmt} is the first Chinese X-ray telescope, and is equipped with three slat-collimated instruments: the Low Energy X-ray telescope \citep[LE, 1--12 keV; ][]{hxmt-le}, the Medium Energy X-ray telescope \citep[ME, 5--35 keV; ][]{hxmt-me} and the High Energy X-ray telescope \citep[HE, 20--350 keV; ][]{hxmt-he}, providing capabilities for broadband X-ray timing and spectroscopy. \hxmt\ carried out high-cadence observations of \psrtar\ starting on MJD 59738.17,  $\sim$0.5 days after the detection by MAXI. The set of 79 ME and 77 HE observations includes runs P0404275001 -- P0404275024.
The HE and ME data were studied to investigate the timing and spectral properties of the source because of their broadband X-ray coverage and good time resolution ( $\sim$2~$\mu$s for HE, $\sim$20~$\mu$s for ME).  We analyzed  the data using the \hxmt\
Data Analysis Software  (HXMTDAS) version 2.05. The ME and HE data were calibrated  by using the scripts  {\tt mepical} and {\tt hepical}, respectively. The good time intervals were individually selected from the scripts {\tt megtigen} and {\tt hegtigen} with the standard criteria, that is, ELV $>$ 0, the satellite located outside the South Atlantic Anomaly region and the offset angle from the pointing direction is smaller than $0\farcs04$. 
Totally, 73 type I X-ray bursts have been observed by ME \citep{Chen2022}, which were removed in the timing and spectral analysis. No bursts were found in HE because the burst spectra are usually soft. Background subtracted light curves for the ME and HE were generated (see Figure~\ref{fig:outburst}). The outburst light curve of \hxmt\ ME showed a fast rise during the early days of the outburst and a slow decay in the next 25~d. However, the \hxmt\ HE light curve for the 20--50 keV band dropped continuously from $\sim100$ cts\,s$^{-1}$ since the start of  the observations to the quiescent level during the outburst.  The spectra and their response matrix files are produced by the tools {\tt hespecgen} and {\tt herspgen} for HE, and {\tt mespecgen} and {\tt merspgen} for ME, respectively. Finally, we obtained the cleaned events using {\tt mescreen} and {\tt hescreen} and  barycentered with the tool {\tt hxbary}.

\subsection{\nicer}\label{sec:nicer}

\begin{figure}
%\vspace{-0.9cm}
\includegraphics[width=8.5cm]{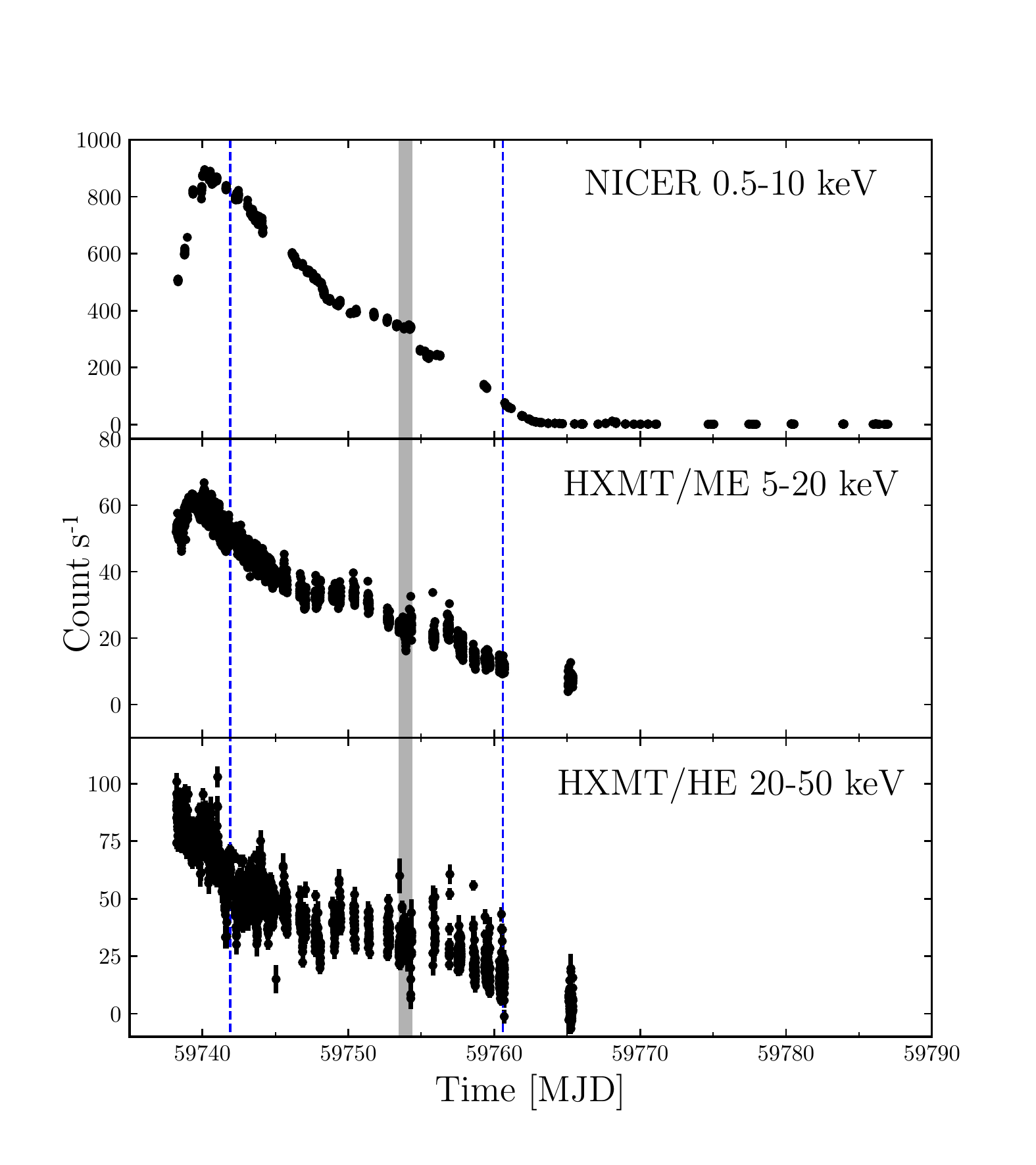}
The light curves of \psrtar\ during its 2022 outburst. From top to bottom: the 100 s binned light curves from \nicer, \hxmt\ ME, and \hxmt\ HE are displayed, respectively. The energy range of each light curve is indicated in each panel. The grey area marks the time of \nustar\ observation. Time intervals covering X-ray bursts are removed. The blue dashed lines represent the time interval that we used for the derivation of our timing models, see Table~\ref{table:eph} and Sect.~\ref{sec:timing}.
\label{fig:outburst}
\end{figure}

\nicer\ started regular observations of \psrtar\ on 2022 June 7 (MJD 59737.6; obs. ID 5202820101), several hours after the detection by MAXI, and ended these on 2022 August 3 (MJD 59794.01; obs. ID 5533013504) when the source was already several weeks in its quiescent state.  We performed the standard data processing using the NICER Data Analysis Software (NICERDAS). The default filtering criteria were applied to extract the cleaned event data. Next, we extracted 1-s light curves by {\tt xselect} for the 0.5--10 and 12--15 keV bands to search for type I X-ray bursts and flaring background, respectively.  Time intervals containing X-ray bursts or background flares were removed in further analysis. The 0.5--10 keV light curve showed a profile similar to \hxmt\ ME, i.e., a fast rise in first few days, and a decay to quiescent level during the next $\sim25$ days (see the top panel of Figure~\ref{fig:outburst}). 

The background spectra were produced from the {\tt nibackgen3C50} tool \citep{Remillard22}. The redistribution matrix file and ancillary response file were generated from the tools {\tt nicerarf} and {\tt nicerrmf}, respectively. 

In the pulse-profile analysis of the \nicer\ data (see Sect.~\ref{sec:pulse_profile}) we used obs. IDs 5533013001--5533013502 (MJD 59769.566--59786.993; cleaned exposure time 14.388 ks) performed when the source had already reached its quiescent state with undetectable (pulsed) flux levels, for the estimation of the local background to obtain background corrected fractional amplitudes.  

\subsection{\nustar}\label{sec:nustar}

\nustar\ observed \psrtar\ on 2022 June 23  (obs. ID 90801315001; MJD $\sim 59753.456-59754.413$) for about 35.7 ks. We cleaned the event file using the  \nustar\ pipeline tool {\tt nupipeline} for both FPMA and FPMB. The light curves and spectra were extracted from a circular region with a radius of  100\arcsec\ centered on the source location by using {\tt nuproducts}, and the response and ancillary response files were produced simultaneously. To extract the background spectra, we chose a source-free circular background region centered at $(\alpha_{2000},\delta_{2000})=(18^{\rm{\scriptsize{h}}}16^{\rm{\scriptsize{m}}}28\fs4048,-19\degr39\arcmin56\farcs308)$ with a radial aperture of $100\arcsec$. From the light curves, four type I X-ray bursts were identified \citep[see also][]{Mandal23}. For the timing and spectral analysis the time intervals during these bursts are discarded resulting in a total exposure time of 29.6~ks. 

In the timing analysis, we barycentered event data from the source region, adopting a $90\arcsec$ extraction radius, using {\tt HEASOFT} multi-mission tool {\tt barycorr v2.16,} with \nustar\ fine clock-correction file \#142, yielding time tags accurate at 60--100~$\mu$s level \citep[][]{Bachetti2021} in absolute time. To obtain background corrected timing characteristics, such as fractional amplitudes, we used the above mentioned background region, however, now with a 90\arcsec\ extraction radius.

%%%%%%%%%%%%%%%%%%%%%%%%%%%%%%%%%%%%%%%%%%%%%%%%%%%%%%%%%%%%%%%%%%%%%%%%%%%%%%%%%%%%%%%%%

\section{Timing analysis}
\label{sec:timing}

Irrespective the instrument in timing analyses we converted the Terrestial Time (TT) arrival times of the selected events to arrival times at the solar system barycenter (in a Barycentric Dynamical Time (TDB) time scale). For this process, throughout in this work we used: (1) the JPL DE405 solar system ephemeris; (2) the instantaneous spacecraft position with respect to the Earth's center;  (3) the (most accurate) VLA position of \psrtar, $\alpha_{\rm 2000} = 18^{\rm h}16^{\rm
m}52\fs41168$ and $\delta_{\rm 2000} = -19\degr37\arcmin57\farcs40138$ \citep{ATel15481}, which differ by $0\farcs4$ in declination from the \swift\ location \citep{ATel15421}. Moreover, in the selection procedures we ignored time intervals during which type I X-ray bursts or high-background flaring occurred.

\subsection{\hxmt\ and \nicer\ Timing analysis}

Initially, we constrained the timing analysis to \hxmt\ HE data using only events with measured energies between 20 and 35 keV. Before generating phase-coherent timing models (ephemerides) describing accurately the spin frequency evolution across the outburst we first need to correct the barycentered event arrival times for the periodic orbital motion effects. Early mission data covering only a part (MJD 59738--59752) of the full dataset available for \hxmt\ HE analysis had been used to improve the orbital parameters of \psrtar\ applying an optimisation scheme based on a SIMPLEX algorithm \citep[see][for more details]{deFalcob,ZLi21} by iteratively improving the $Z_2^2(\phi)$-test statistics \citep{buccheri1983} with respect to the pulse frequency $\nu$, the orbital period $P_{\rm orb}$, and time of the ascending node $T_{\rm asc}$.
The optimised values for the (most critical) orbital parameters $P_{\rm orb}$ and $T_{\rm asc}$, while fixing the eccentricity to zero and the projected semi-major axis $a_{\rm x} \sin i$ to the value quoted in \citet{ATel15425} are listed in Table \ref{table:eph}. The obtained values are fully consistent within $1\sigma$ with the best orbital
parameters obtained later by \citet{Bult22} using the full \nicer\ database. Therefore, we kept the values for the orbital parameters as derived from our three parameter optimisation scheme using 20--35 keV \hxmt\ HE data. 

%%%%%%%%%%%%%%%%%%%%%%%%%%%%%%%%%%%%%%%%%%%%%%%%%%%%%%%%%%%%%%%%%%%%%%%%%%%%%%%%%%%%%%%
%%%%%%%%%%%%%%%%%%%%%%%%%%%%%%%%%%%%%%%%%%%%%%%%%%%%%%%%%%%%%%%%%%%%%%%%%%%%%%%%%%%%%%%
\begin{table}[t] 
{\small
\caption{Positional, orbital and spin parameters derived in this work and used as fixed values from literature for \psrtar.}
\centering
\begin{tabular}{lcc} 
\hline \hline 
Parameter                      & Values                                                               &Units    \\
\hline 
\noalign{\smallskip}  
$\alpha_{2000}$                & $18^{\hbox{\scriptsize h}} 16^{\hbox{\scriptsize m}} 52\fs41168(12)$ &         \\   
$\delta_{2000}$                & $-19\degr37\arcmin57\farcs40138(473)$                                &         \\             
JPL Ephemeris                  & DE405                                                                &         \\           
$ P_{\rm orb} $                & $17402.5786(60)$                                                       &s        \\             
$ a_{\rm x}\, \sin i$          & $0.262\,948(18)$                                                         &lt-s     \\             
$ e $                          & $0$ (fixed)                                                          &         \\             
$T_{\rm asc} $                & $59738.875\,632(4)$                                                    &MJD (TDB)\\             
\hline
\multicolumn{3}{c}{Constant Frequency model}\\
\hline 
Validity range                 & $59741.9 - 59760.6$                                                   &MJD (TDB)\\                      
$t_0$ (Epoch)                  & $59741.0$                                                            &MJD (TDB)\\
$\nu$                          & $528.611\,105\,832(4)$                                               &Hz       \\
\chiq{}/d.o.f                  & $86.78 / (47-1) = 1.886$                                             &         \\
\hline 
                               & Quadratic / Spin-up model                                                        &         \\
\hline 
Validity range                 & $59741.9 - 59760.6$                                                   &MJD (TDB)\\                      
$t_0$ (Epoch)                  & $59741.0$                                                            &MJD (TDB)\\
$\nu$                          & $528.611\,105\,774(12)$                                              &Hz       \\
$\dot{\nu}$                    & $(9.0\pm2.1)\times10^{-14}$                                            &$\rm{Hz~s^{-1}}$ \\
\chiq{}/d.o.f                  & $68.18 / (47-2) = 1.515$                                             & \\
\noalign{\smallskip}  
\hline  
\end{tabular}

\label{table:eph} 
}
\end{table}

%%%%%%%%%%%%%%%%%%%%%%%%%%%%%%%%%%%%%%%%%%%%%%%%%%%%%%%%%%%%%%%%%%%%%%%%%%%%%%%%%%%%%%%
%%%%%%%%%%%%%%%%%%%%%%%%%%%%%%%%%%%%%%%%%%%%%%%%%%%%%%%%%%%%%%%%%%%%%%%%%%%%%%%%%%%%%%%

After correcting for the orbital motion effects we performed a time-of-arrival \citep[ToA; see][for more details]{Kuiper2009} analysis to obtain accurate estimates for the pulse frequency $\nu$, and (if required) its first time derivative $\dot{\nu}$ over a certain time interval. We obtained 69 ToA's for \hxmt\ HE (20--35 keV) covering MJD 59738--59761. 
Similar analyses yielded 73 ToA's for \hxmt\ ME restricted to the 10--20 keV band (MJD 59738--59761), and 26 ToA's for \nicer\ using 1.8--10 keV events across MJD 59737--59761. 

\begin{figure} 
\centering
\includegraphics[width=0.9\linewidth]{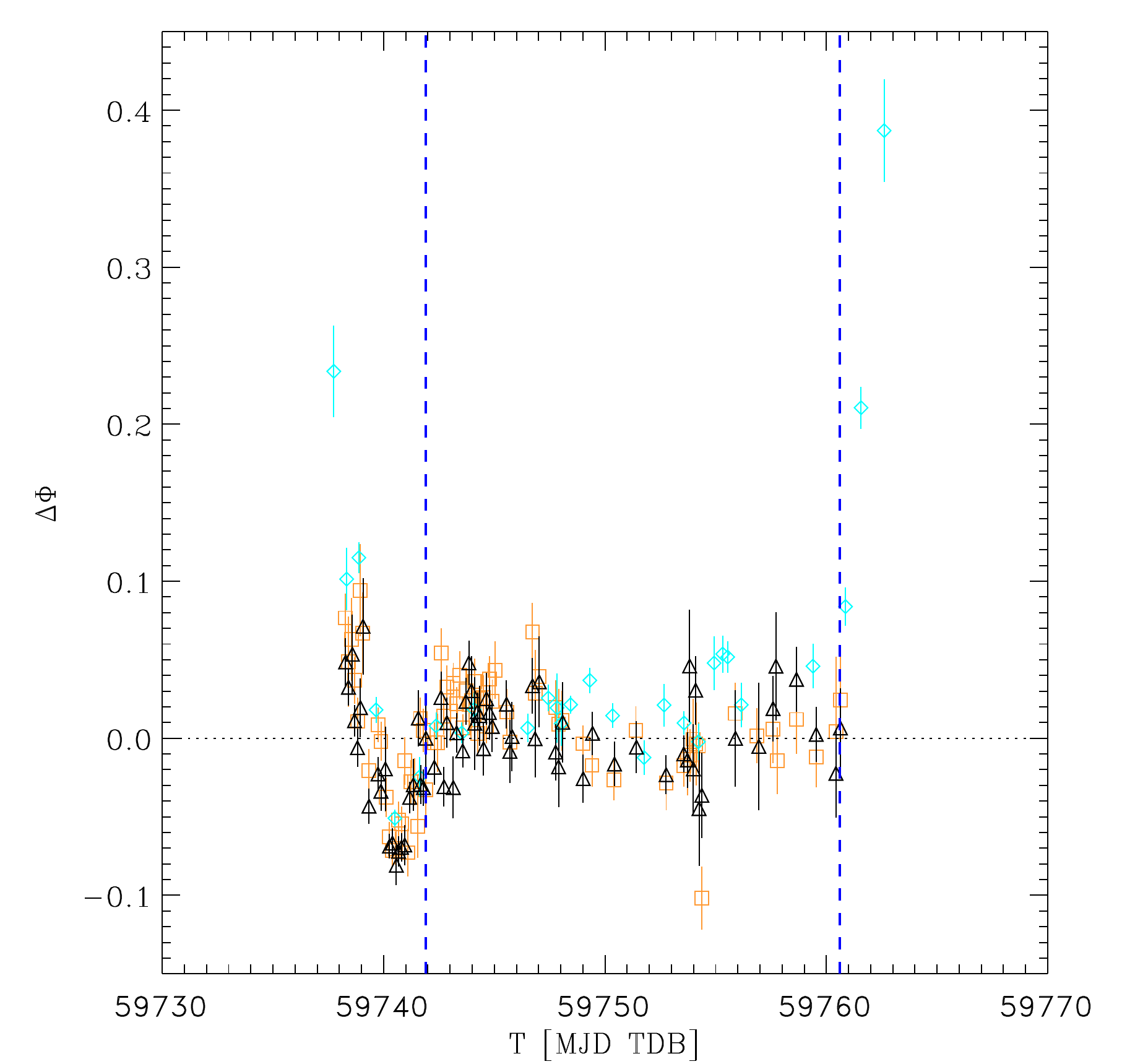}
\caption{Pulse phase residuals of \hxmt\ HE (open black triangles; 20--35 keV), \hxmt\ ME (open orange squares; 10--20 keV) and \nicer\ (open aqua diamonds; 1.8--10 keV) ToAs with respect to a two-parameter timing model (including a quadratic / spin-up term) valid for the range MJD 59741.9--59760.6. These models are based on solely \hxmt\ HE ToAs (black data points). The interval bounded by the two blue dashed lines represents this time interval that has been used to derive the timing models given in Table \ref{table:eph}. Note, that there are no ToA measurements beyond MJD 59760.6 for \hxmt\ HE (and ME) contrary to \nicer.
}
\label{fig:residuals}
\end{figure}

Fitting the complete \hxmt\ HE ToA dataset with a constant frequency model resulted in a very poor fit with an unacceptable \chiq{}. Ignoring 22 ToA's from the first $\sim 5$ days of the outburst yielded
reasonable fits assuming a constant frequency model and a model including a frequency derivative term.
Both are listed in Table \ref{table:eph}. The constant frequency model is fully consistent at $1\sigma$ confidence with the corresponding model, based on solely \nicer\ data, shown in Table 1 of \citet{Bult22} for 46 degrees of freedom. 

The model with both the $\nu$ and $\dot{\nu}$ terms, however, improves upon the constant frequency model by reducing the \chiq{} from 86.78 to 68.18 adding one fit parameter. Applying a maximum likelihood ratio test to assess the model improvement yields an improvement of $(86.78-68.18)^{0.5}=4.3\sigma$, adopting one degree of freedom (=the additional fit parameter).\footnote{The square-root relation holds only for the case of one additional fit parameter (= one degree of freedom)}
The measured $\dot{\nu}$ of $(9.0\pm2.1)\times10^{-14}~\rm{Hz~s^{-1}}$ can be interpreted as a spin-up across the time interval MJD 59741.9--59760.6.
We have also investigated the effects of employing a flux-bias model in the ToA fit
procedure by adding a (flux) bias term in the form,
$\Phi_{\rm{\scriptsize{bias}}}=\alpha (C/C_{\max})^{\Gamma}$ with $C_{\max}=895.37\ {\rm c\,s^{-1}}$ 
(the maximum observed count rate in the 0.5--10 keV NICER bandpass), to the pulse-phase 
residual relation in analogy with \citet{Bult22} (see their Eq. 1). Note, that in Table 1 of 
\citet{Bult22} both optimized flux-bias model parameters, their $b$ and $\Gamma$, are listed 
abusively with a plus sign. Taken this sign inconsistency into account we could reproduce 
their findings and obtained $\Gamma=-1.05 \pm 0.15$, consistent within one sigma confidence 
with their value of $-1.20 \pm 0.20$ employing the {\em{full}} NICER ToA dataset running 
from MJD 59737.748--59762.610.
Constraining the NICER ToA dataset to MJD 59742.370--59762.610, and so ignoring the rising
and maximum part of the light curve (see Figure~\ref{fig:outburst}), we obtained a flux-bias index of 
$\Gamma=-0.77 \pm 0.09$, however, as with the full range NICER model shown before, with a 
statistically unacceptable fit quality of $\chi^2_r \sim 3.54$ compatible with the results
shown in \citet{Bult22}.
Next, we fitted a constant frequency- plus flux-bias model with a {\em fixed} index $\Gamma=-0.77$
to the \hxmt-HE ToA dataset covering MJD 59741.9--59760.6, the decaying part of the outburst. This model improves upon the constant frequency model by reducing the $\chi^2$ from
86.78 to 83.10 adding one free parameter (the flux-bias model scale $\alpha$), which represents a 
$(86.78-83.10)^{0.5}=1.92\sigma$ improvement with respect to the constant frequency model, applying
a maximum likelihood ratio-test. The best fit flux-bias scale parameter is $\alpha=(-7.1\pm 3.7)\times 10^{-3}$, while $\nu=528.611\,105\,851(11)$ Hz.
Therefore, the spin-up model for the decaying part of the outburst (covering MJD 59741.9--59760.6) provides a better data description than the flux-bias model for the \hxmt-HE ToA dataset.

The pulse-phase residuals of the complete sets of ToA's from all three instruments combined (each determined for a different energy band) after folding upon the (\hxmt\ HE 20--35 keV based) spin-up model are shown in Figure \ref{fig:residuals}.  
Irrespective the energy band each instrument showed a quite similar behaviour: during the first $\sim 5$ days of the outburst - the rising flux part -there is a structured residual feature which can be described by a (short-duration) spin-down model. This feature is the cause of the very poor \chiq{} when the full ToA datasets are considered \citep[see also Table 1 of ][indicating unacceptable \chiq{} fit values]{Bult22}. Also, the steep positive excursions for the \nicer\ ToAs at the early beginning and very end (not covered by \hxmt\ observations) of the outburst are striking in Figure~\ref{fig:residuals}. It is clear that even a (bolometric) flux bias model as well as the constant frequency and quadratic models considered in this work cannot describe the {\em{full}} set of residuals, consistent with \citet{Bult22}. 
 
\begin{figure*}[t]
\centering
\includegraphics[width=18cm]{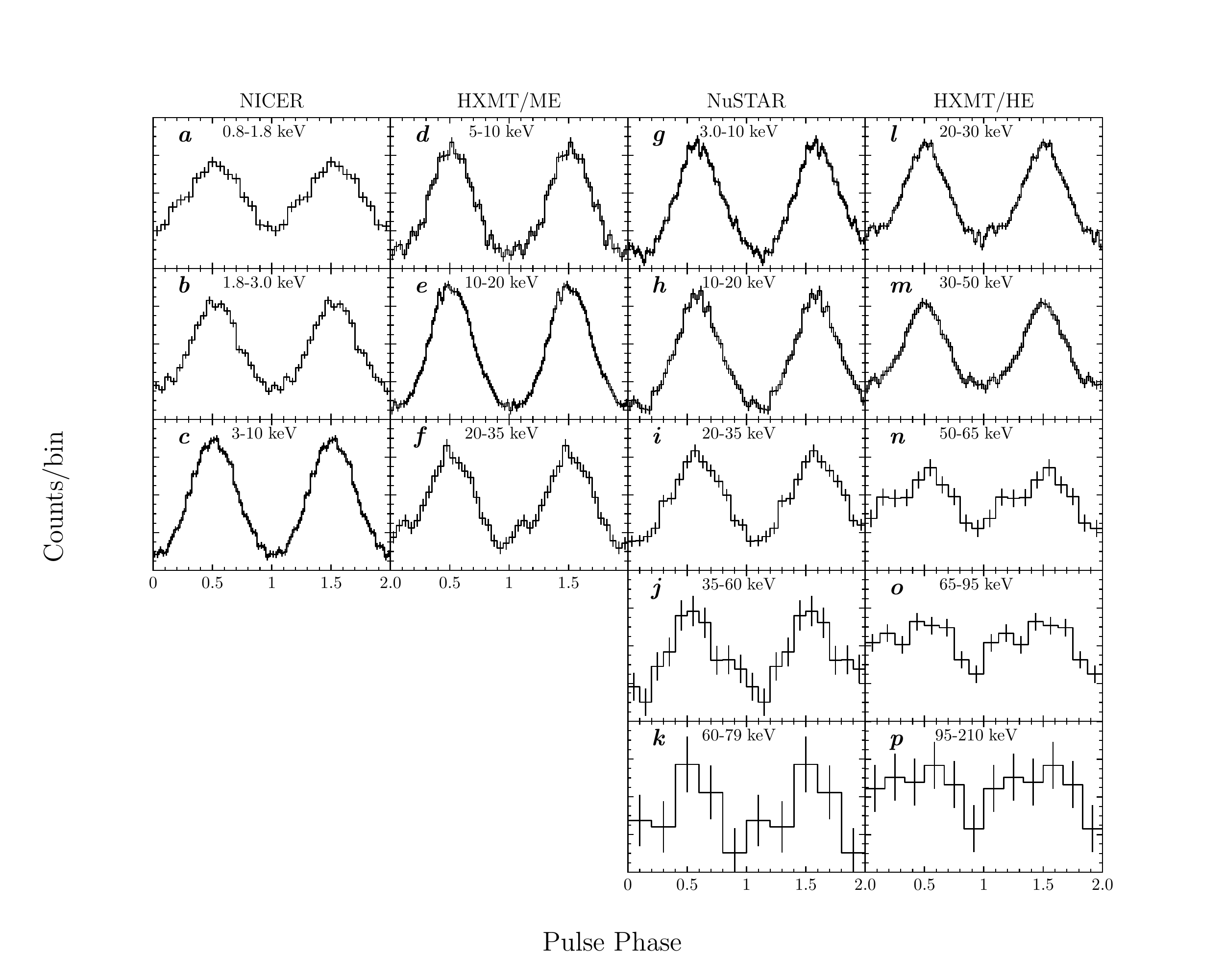}
\caption{The 0.8--210  keV broadband pulse-phase distributions of \psrtar\ observed by \nicer\ (panels a--c, 0.8--10 keV), \nustar\ (panels g--k, 3--79 keV), and \hxmt\ (panels d--f, 5--35 keV for ME; panels l--p, 20--210 keV for HE). 
In order to improve clarity, two cycles are shown.  The error bars represent $1\sigma$ errors. The morphology is almost  unchanged with energy. All profiles reach their maximum near phase $\sim$0.5.}
\label{pulse_profile}
\end{figure*}

%%%%%%%%%%%%%%%%%%%%%%%% Fig-4 %%%%%%%%%%%%%%%%%%%%%%%%
\begin{figure*} %[t]
  \centering
  \includegraphics[width=8.3cm]{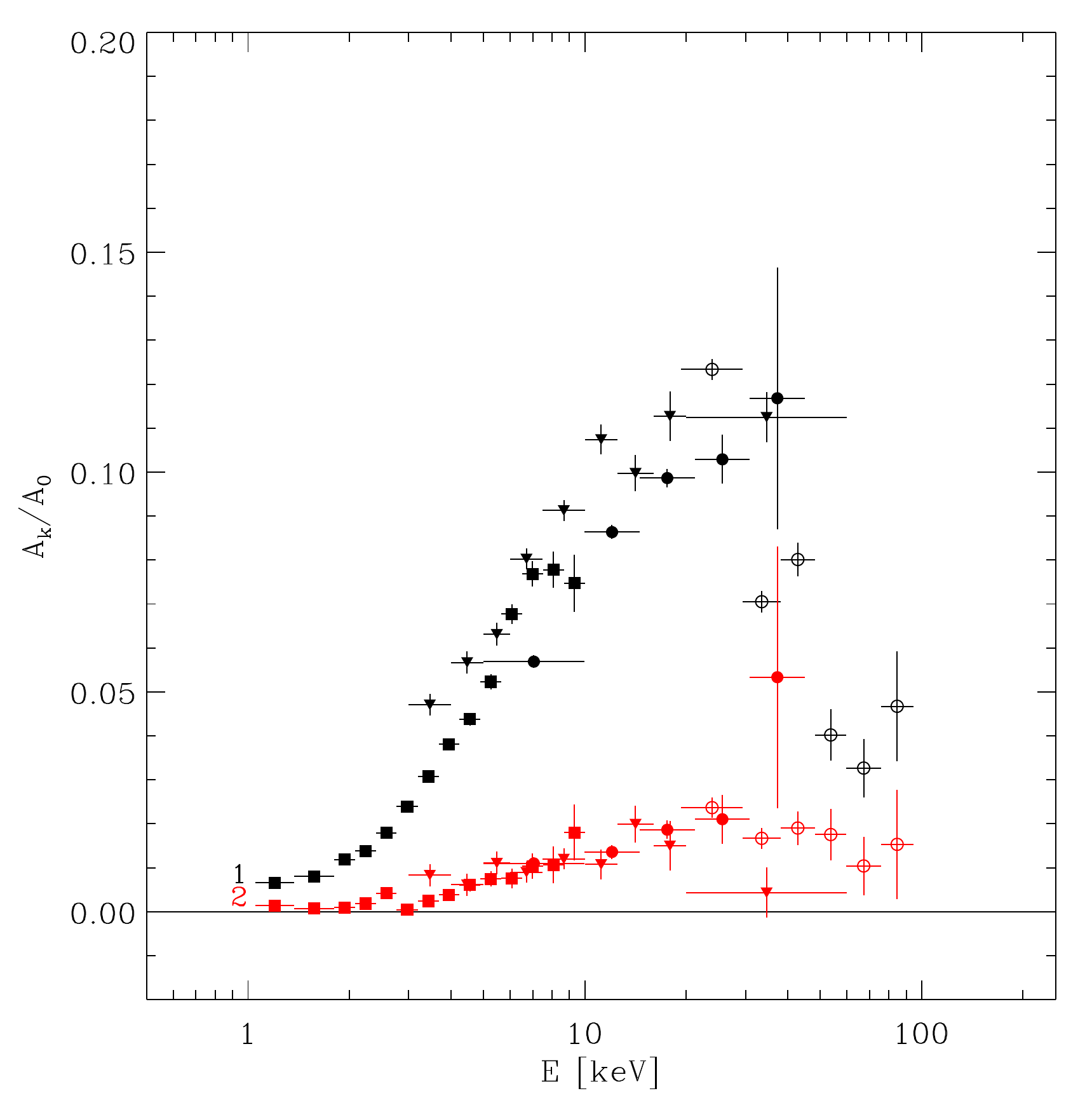}
  \hspace{0.5cm}
  \includegraphics[width=8.5cm]{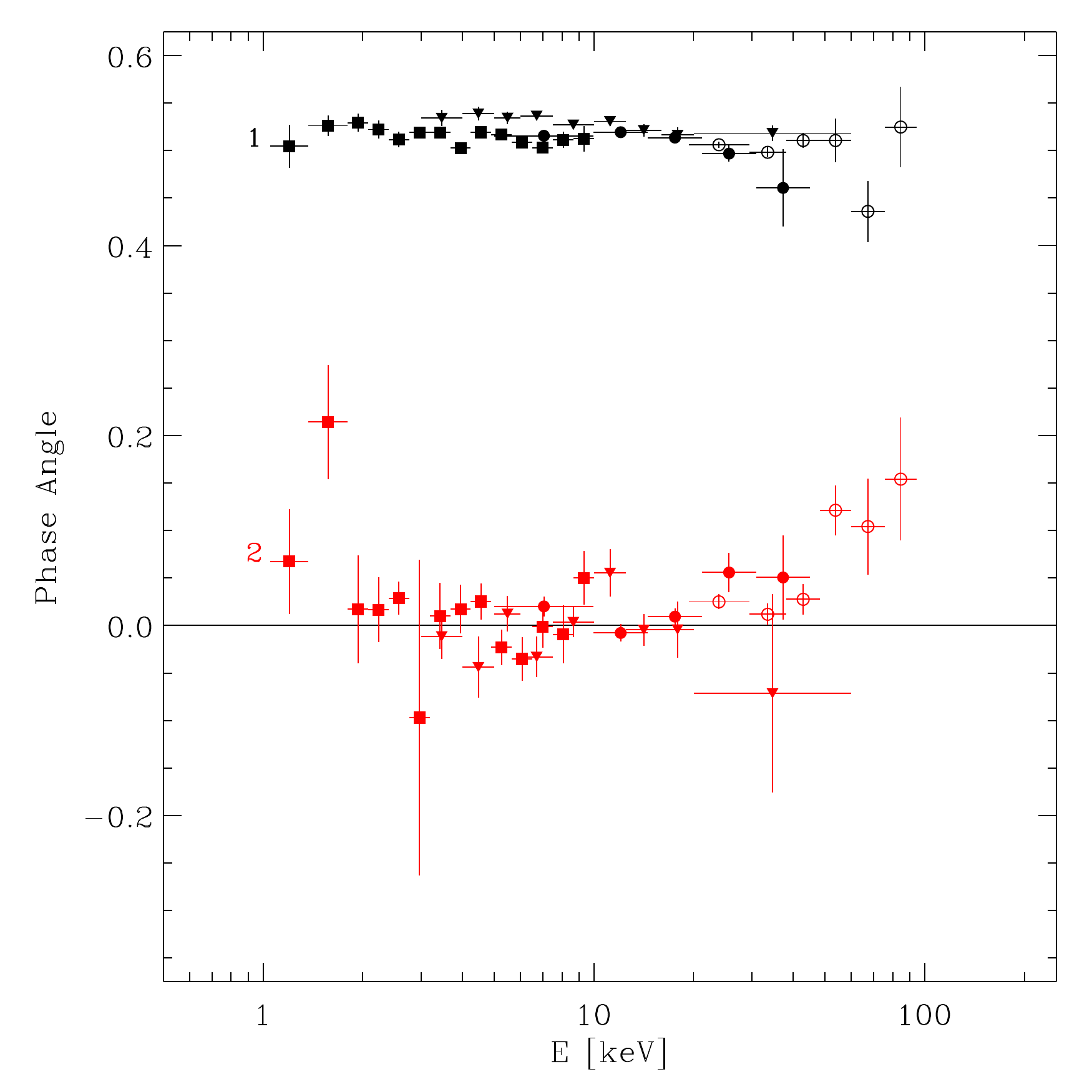}
\caption{Left: The fractional amplitude of various harmonics: 1 -- fundamental (black crosses) and 2 -- first overtone (red crosses) as a function of energy using background-corrected data from \nicer\ (filled squares), \nustar\ (triangles), \hxmt\ ME (filled circles) and \hxmt\ HE (open circles).
Right: The phase angle (in units radians/$2\pi$) as a function of energy for the two Fourier components, the fundamental (in black) and the first overtone (in red), which are independent of the background subtraction method.}
  \label{fig:phase_angle}
\end{figure*}
%%%%%%%%%%%%%%%%%%%%%%%% Fig-4 %%%%%%%%%%%%%%%%%%%%%%%%

\subsection{Pulse profile analysis}
\label{sec:pulse_profile}

With the accurate orbital and (spin-up) timing models (see Table \ref{table:eph}) we converted the event arrival times from each high-energy instrument to pulse phases. To obtain high-statistics pulse-phase distributions (pulse profiles including background) we combined the phase information for each instrument across the validity interval (MJD 59741.9--59760.9) and sorted these on (measured) energy. We verified for each instrument that the pulse shape for each individual observation constituting the combination remained the same within statistical uncertainties.
The measured pulse-phase distributions  for the four different high-energy instruments are shown in Figure~\ref{pulse_profile} for various different energy bands covering the 0.8--210~keV range. 

Panels (a)--(c) of Figure~\ref{pulse_profile} show the NICER pulse profiles for the energy intervals 0.8--1.8, 1.8--3, and 3--10 keV, respectively, while panels (d)--(f) show the \hxmt\ ME profiles for the 5--10, 10--20 and 20--35 keV bands.
The \nustar\ profiles are displayed in panels (g)--(k) covering the 3--79 keV energy range. 
Noteworthy is that comparing the \nicer\ and \nustar\ 3--10 keV distributions, shown in panels (c) and (g), respectively, a small phase shift of $0.065 \pm 0.020$ is detectable which is just compatible with the absolute timing accuracy of \nustar\ \citep{Bachetti2021}. Pulsed emission is detected significantly up to $\sim$60 keV, with a non-uniformity significance of $5.4\sigma$ for the 35--60 keV band (panel j) adopting a $Z_2^2$-test \citep{buccheri1983}.
This test applied to the 60--79 keV band (panel k) yielded an insignificant result of $1.3\sigma$.
Finally, panels (l)--(p) show the \hxmt\ HE profiles for energies of 20--30, 30--50, 50--65, 65--95 and 95--210 keV, respectively.
Interestingly, pulsed emission is detected significantly up to $\sim$95~keV. The 65--95 keV band (panel o) shows a non-uniform significance of $4.4\sigma$ applying a $Z_2^2$-test with indications for a second weaker pulse near phase 0.15 that seems to pop-up at energies above $\sim$50~keV (panels n--o). The non-uniformity significance for the distribution shown in panel (p) for the 95--210 keV band is about $1.1\sigma$, and so insignificant.

To obtain quantitative information about morphology changes of the pulse profiles as a function of energy, we fit the pulse profiles with a truncated Fourier series given by a formula
%\begin{equation}
%F(\phi)=A_{1}\sin[2\pi(\phi-\phi_1)]+A_{2}\sin[4\pi(\phi-\phi_2)]+B,
%\label{eq10}
%\end{equation}
\begin{equation}
F(\phi) = A_0 + \sum_{k=1}^2 A_k \ \cos[2\pi\ k (\phi-\phi_k)],
\label{eq10}
\end{equation}
where $A_{1}$ and $A_{2}$ are the amplitudes,  $\phi_1$ and $\phi_2$ are the phase angles (in radians/$2\pi$), of the fundamental and the first overtone, respectively, and $A_0$ is the constant level of the profile. We verified that higher harmonics are not required to accurately describe the data.
The left panel of Figure \ref{fig:phase_angle} shows the background corrected fractional amplitudes $A_k/A_0$ of the fundamental (black) and first overtone (red) for all four instruments covering the 0.8--95 keV band, across which we detected significant pulsed emission.

The background correction determination was straightforward for \nicer\ and \nustar. We used for the former non-imaging instrument the data from observations when the source was completely Off (see Sect. \ref{sec:nicer}) with a combined Good Time Interval (GTI) exposure of 14.388 ks. Along with the On-time GTI exposure of 51.735 ks this yielded a scale factor $s_{\scriptsize{\rm{bg}}}$ of $3.5957$ to be applied in the background subtraction.
For the latter (imaging) instrument the data from a different source-free location in the field-of-view with equal area as source region ($s_{\scriptsize{\rm{bg}}}\equiv 1.0$) was used. 

For the non-imaging \hxmt\ ME/HE (with GTI exposures of 186.171 ks and 132.093 ks, respectively, for the validity interval during the On source period) the last observations of \psrtar\ performed on MJD 59765 (2022 July 5) totaling exposure times of 13.474 ks and 9.022 ks for ME and HE, respectively, were used to estimate the underlying background, when the source did not show pulsed emission anymore, but was not completely Off \citep[see, e.g.,][who reported the detection of the source at a $4.1\sigma$ level in 5.2 ks \swift\ XRT data taken a day later on MJD 59766]{ATel15506}. 

Thus, the ME/HE event data of MJD 59765 still contain a tiny contribution from \psrtar, that would yield an over-subtraction when applied uncorrected in the background subtraction procedure. Demanding that the background subtraction did not yield a negative value for energy bands at the high end of the sensitivity window of both instruments (ME $\gtrsim 30$ keV; HE $\gtrsim 95$ keV) we could estimate the genuine background contribution in both samples. For the ME/HE, we found that $\sim$9\% and $\sim$5\% of the events from the background sample was originating from the `almost' Off \psrtar, respectively.
Taking these fractions into account in the background subtraction process showed that the $A_1/A_0$ (and also $A_2/A_0$) values obtained for ME/HE  matched with \nustar\ values obtained in the overlapping energy range.

It is clear from the left panel of Figure \ref{fig:phase_angle} that the pulsed fraction of the fundamental (black) gradually increases with energy from $\sim$0.5\% to 10\%--12\% going from $\sim$0.8 to $\sim$10~keV, where a plateau is reached. Above $\sim$35~keV a decrease sets in to values of 3\%--5\% above 50 keV.
The first overtone (red) saturates at $\gtrsim$10~keV to a $\sim$2\% value, without a clear decrease beyond. This means that the $A_2/A_1$ ratio increases the higher the energy, implying that the first overtone contributes more significantly at higher energies.

The right panel of Figure \ref{fig:phase_angle} shows the phases of the maxima of the fundamental (black) and first overtone (red) for the four instruments, specifying the alignment of the components constituting the pulse profile. It is noteworthy that these quantities, $\phi_1$ and $\phi_2$, are independent on the used background method.

For \nustar, which provided a snapshot during the full outburst, we have subtracted 0.065 (in phase units) -still within its absolute timing accuracy - from the measured $\phi_1$ and $\phi_2$ values for alignment purposes with \nicer. It is clear that irrespective the instrument both quantities are stable within statistical uncertainties till $\sim$50 keV, beyond which the first overtone moves closer to the fundamental. Combined with the increase of the $A_2/A_1$ ratio at those energies this means that a second bump/maximum in the pulse profile in front of the main maximum becomes  more prominent at $\gtrsim$50~keV energies.

%We also studied the  \hxmt/ME/HE pulse profile evolution over  the outburst. Generally, the amplitudes of the fundamental and first overtone were almost unchanged during the outburst. The phase angle of the first overtone first increased and then decreased during the first five days of the outbursts, and remained constant in the next 15 days. The phase angle of the fundamental shifted less than 0.05 cycles.

%\begin{figure}
% \includegraphics[width=8cm]{pro_en_fit.eps}
% \includegraphics[width=8cm]{pro_Ar_fit.eps}
%\includegraphics[width=9cm]{phi_ratio.eps}
%\caption{Upper panel, the amplitude ratio, $A_2/A_1$ of two Fourier components. Lower panel, the phase of two components. The black and red data present the phases $\phi_1$ and $\phi_2$, respectively. Open square, filled square, open circle and filled circle are from \nicer, \nustar, \hxmt\ ME and \hxmt\ HE observations, respectively. }
%\label{fig:phase_angle}
%\end{figure}

%\begin{figure*}
%\includegraphics[width=9cm]{HEpro_phi_fit.eps}
%\includegraphics[width=8cm]{HEpro_tim_fit.eps}
%\caption{Left upper panel, the amplitude ratio of two components over time. 
%Left lower panel, the phase of two components. Green and magenta triangle represent
%the fundamental and first overtone of ME, respectively. Red and blue square represent
%the fundamental and first overtone of HE.
%Right panel, from bottom to up, the 12 \hxmt/HE pulse profiles in 30--208 keV
%represent the time intervals marked in left panel. }
%\end{figure*}

%%%%%%%%%%%%%%%%%%%%%%%%%%%%%%%%%%%%%%%%%%%%%%%%%%%%%%%%%%%%%%%%%%%%%%%%%%%%%%%%%%%%%%%%%

\section{Spectral analysis}
\label{sec:spec}

The spectral analysis was carried out using \textsc{xspec} version 12.12.1 \citep{arnaud96}.  We first performed the joint spectral evolution analysis for \nicer\   and \hxmt\  observations (Sect.~\ref{sec:nicer_spec}).  And then, we studied the quasi-simultaneous broadband spectra of \nicer, \nustar\ and \hxmt\ (Sect.~\ref{sec:broad_band}). All spectra are grouped by the tool {\tt ftgrouppha}.  For \nicer\ and \nustar\ spectra, we applied the optimal binning method \citep{Kaastra2016}. The \hxmt\ spectra were grouped with a minimal signal-to-noise ratio of 3 related to their backgrounds. All uncertainties of the spectral parameters are provided at a $1\sigma$ confidence level for a single parameter. 

% as suggested by the instrumental team.\footnote{\url{https://heasarc.gsfc.nasa.gov/docs/nicer/analysis_threads/spectrum-grouping/}}

\subsection{The spectral evolution of joint \nicer\ and \hxmt\ observations}
\label{sec:nicer_spec}

There were many quasi-simultaneous \nicer\ and \hxmt\  observations of \psrtar, which allows us to perform a detailed broadband spectral analysis for the whole outburst.  We found that each \nicer\ observation in the MJD 59738--59761 range can be (partially) covered by 1-6 \hxmt\ observations. We combined the \hxmt\ spectra, and then carried out joint \nicer\ and \hxmt\ spectral fitting.  We adopted the energy ranges of 1--10 keV for \nicer\ spectra, 10--35 keV for \hxmt\ ME and 25--150 keV for \hxmt\ HE. We fit all the spectra by using the combination of the thermal Comptonization model, {\tt compps}, in the slab geometry \citep{Poutanen96}, the disk blackbody, {\tt diskbb}, modified by the interstellar absorption described by the model, {\tt tbabs}. This model has been used to fit the broadband spectra of many AXMPs \citep[see, e.g.,][]{gp05,falanga05,falanga05b,falanga11,falanga12,deFalcoa,deFalcob,Kuiper20,ZLI2018,ZLi21}. To account for cross-calibration uncertainties between different instruments a multiplication factor is included in the fits. The factor is fixed at unity for \nicer\ and set free for \hxmt\ ME/HE. The main parameters  are the Thomson optical depth across the slab, $\tau_{\rm T}$, the electron temperature, $kT_{\rm e}$, the temperature of the soft seed photons, $kT_{\rm bb}$, the normalization factor for the seed black body photons, $K_{\rm bb} = (R_{\rm km} /d_{10} )^2$ (with $d_{10}$ being distance in units of 10 kpc), and the inclination angle $\theta$ (fixed at 60\degr) between the slab normal and the line of sight to the observer for the {\tt compps} model; the disk temperature, $kT_{\rm diskbb}$ and  normalization for the multi-black body model, and the hydrogen column density, $N_{\rm H}$. Most of \nicer\ spectra showed emission lines around $\sim$1.6 and $\sim$6.5~keV, therefore, we added two Gaussian components to account for these. The full model is {\tt tbabs (gaussian+gaussian+diskbb+compps)} in \textsc{xspec}. The $\sim$1.6~keV component could have an instrumental origin, while the $\sim$6.5~keV Gaussian component may come from the iron K$\alpha$ line emission.

% The thermal Comptonization model has been used in many other AMXPs. 

% We adopted the phenomenological model, \texttt{tbabs(powerlaw+diskbb)}, to fit each \nicer\ spectrum in 0.5--10 keV individually. For some spectra, there are features in the residuals around $\sim0.5$ keV,  $\sim1.6$ keV and $\sim6.5$ keV.  Therefore, we added 1--3 Gaussian components to account for them. 

% We applied the cutoff power law, {\tt cutoffpl}, plus multi-color disk blackbody,  {\tt diskbb},  modified by the interstellar absorption described by the model, {\tt tbabs}, to fit the joint \nicer\ and \hxmt/ME/HE spectra. The multiplication factor is included to account for the cross-calibration uncertainties of different instruments. The factor is fixed at 1 for \nicer\ and set free for \hxmt.  The main parameters of the model are the power law photon index, $\Gamma$, the e-folding energy of exponential rolloff, $E_{\rm cut}$, the normalization, the disk temperature, $kT_{\rm diskbb}$, and its normalization, and the hydrogen column density, $N_{\rm H}$. Most of \nicer\ spectra showed emission lines around  $\sim1.6$ keV and $\sim6.5$ keV, therefore, we added two Gaussian components to account for them. The full model is {\tt tbabs(gaussian + gaussian + diskbb + cutoffpl)} in \textsc{xspec}. The $\sim1.6$ keV component could be origin from instrument. The  $\sim6.5$ keV Gaussian components may produce from iron K$\alpha$ line. %However, the $\sim0.5$ keV could be also from the instrument.

%

\begin{figure}
\includegraphics[width=8.5cm]{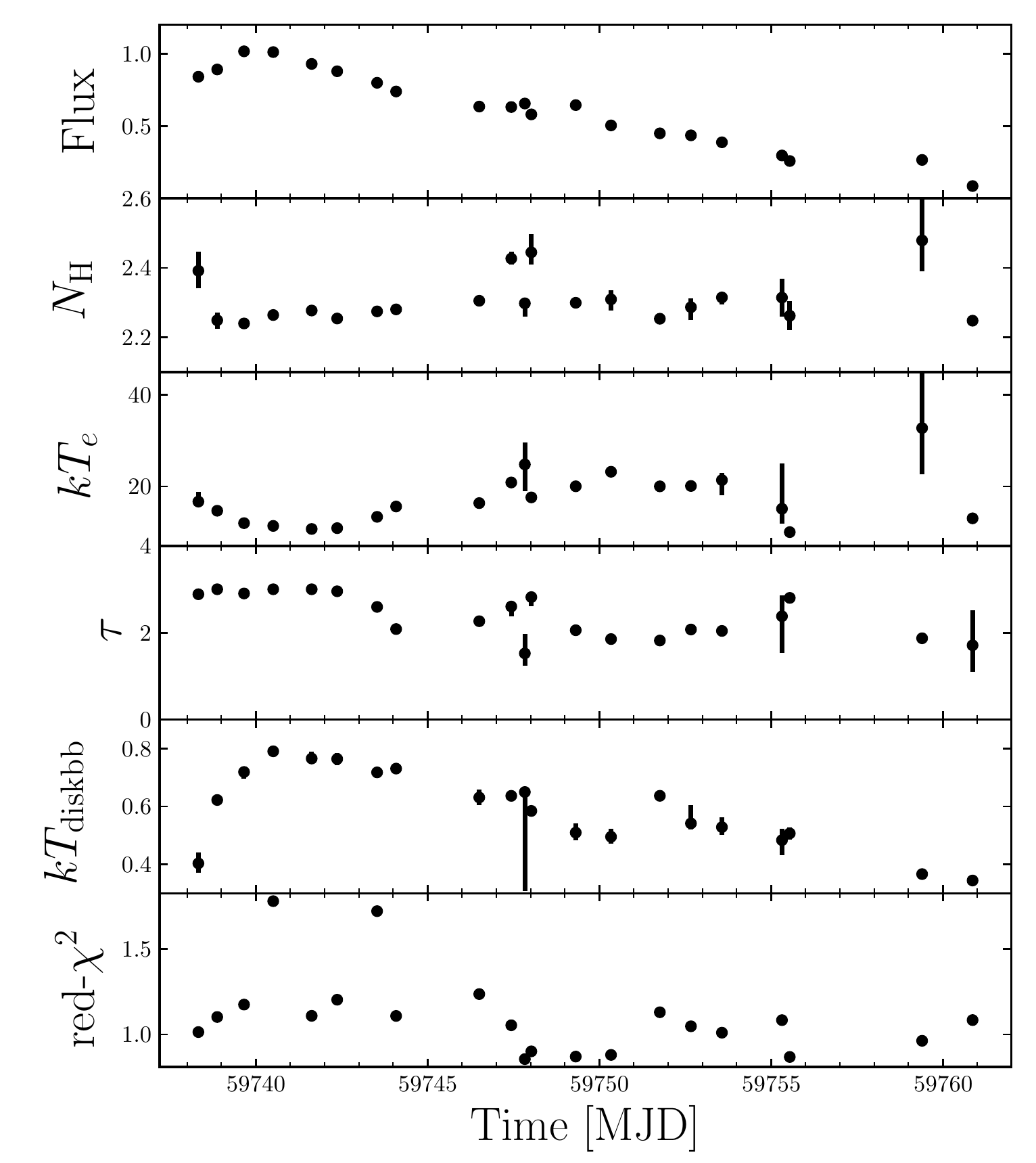}
\caption{The fitted results of joint \nicer\ and \hxmt\ spectra by the model {\tt constant$\times$tbabs(gaussian + gaussian + diskbb + compps)}. From top to bottom, the bolometric flux (in units of $10^{-8}$\,erg\,cm$^{-2}$\,s$^{-1}$), the hydrogen column density (in $10^{22}$~cm$^{-2}$), the electron temperature (in keV), the optical depth, the temperature of disk blackbody (in keV), and the reduced \chiq{} are shown. }
\label{fig:compps}
\end{figure}

The fitted parameters are shown in Figure~\ref{fig:compps}. Most of the spectra can be well fitted with reduced $\chi^2<1.3$.  For two observations, the reduced $\chi^2$ are larger than 1.5, which show no features in the \nicer\ residuals and possibly caused by the spectral evolution during the long exposures. We calculated the unabsorbed bolometric flux in the 1--250 keV range by using the tool \texttt{cflux}. During the outburst,  the disk blackbody temperature increased from 0.4 to 0.8 keV in the first few days and then dropped back to $\sim$0.4 keV. The  hydrogen column density did not change much, and the mean value is $(2.31\pm0.07)\times10^{22}~{\rm cm^{-2}}$. The optical depth was in the range 1.7--3.0, which explains the visibility of hard X-ray emissions during the whole outburst. The electron temperature was around 10--25 keV. The bolometric flux reached a peak value of $1.02\times10^{-8}~{\rm erg~cm^{-2}~s^{-1}}$ during the first few days, and in the next 20 days, it decayed to $8.5\times10^{-10}~{\rm erg~cm^{-2}~s^{-1}}$ at MJD 59760.9.

We also replaced the {\tt compps} component by a cutoff power-law model. 
This model provided compatible results compared with the {\tt compps} model, i.e., $N_{\rm{H}}$, the temperature and normalization of disk black body component, the {\chiq{}}, and the unabsorbed flux.

% \subsection{\hxmt\ spectra}
% \label{sec:hxmt_spec}
% For each \hxmt\ observation, we fit the spectra in 10--35 keV for ME and 25--100 keV for HE. We applied the cutoff powerlaw model to fit the joint ME and HE spectra. The interstellar absorption is ignored since it is prominent at soft X-ray band. The multiplication factor is included to account for the
% cross-calibration uncertainty between HE and ME. The factor is fixed at 1 for ME and set free for HE.  The main parameters of the model are the power law photon index, $\Gamma$, the e-folding energy of exponential rolloff, $E_{\rm cut}$, and the normalization. 
% The joint spectral fit is acceptable for all red-$\chi^2<1$. For first two days during the rise of the outburst,  the powerlaw index and the cutoff energy decreased from 1.6?? to 1.3 and 40 to 20 keV?, respectively, implying the spectra became harder. During the decay of the outburst, the powerlaw index increased from $\sim1.5$ to 2.0, and the cutoff energy also increased from ??, meaning that the spectra became softer. These results are consistent with \nicer\ observations.

% \begin{figure}
% \includegraphics[width=9cm]{hxmt_spectra.eps}
% \caption{The spectral fitted results of \hxmt\ spectra.}
% \end{figure}

\subsection{Broadband spectra including \nustar\ observation}
\label{sec:broad_band}

% All uncertainties in the spectral parameters are given at a $1\sigma$ confidence level for a single parameter. The spectral analysis were carried out using \textsc{xspec} version 12.12.1 \citep{arnaud96}.

\begin{figure}
\includegraphics[angle=-90,width=8cm]{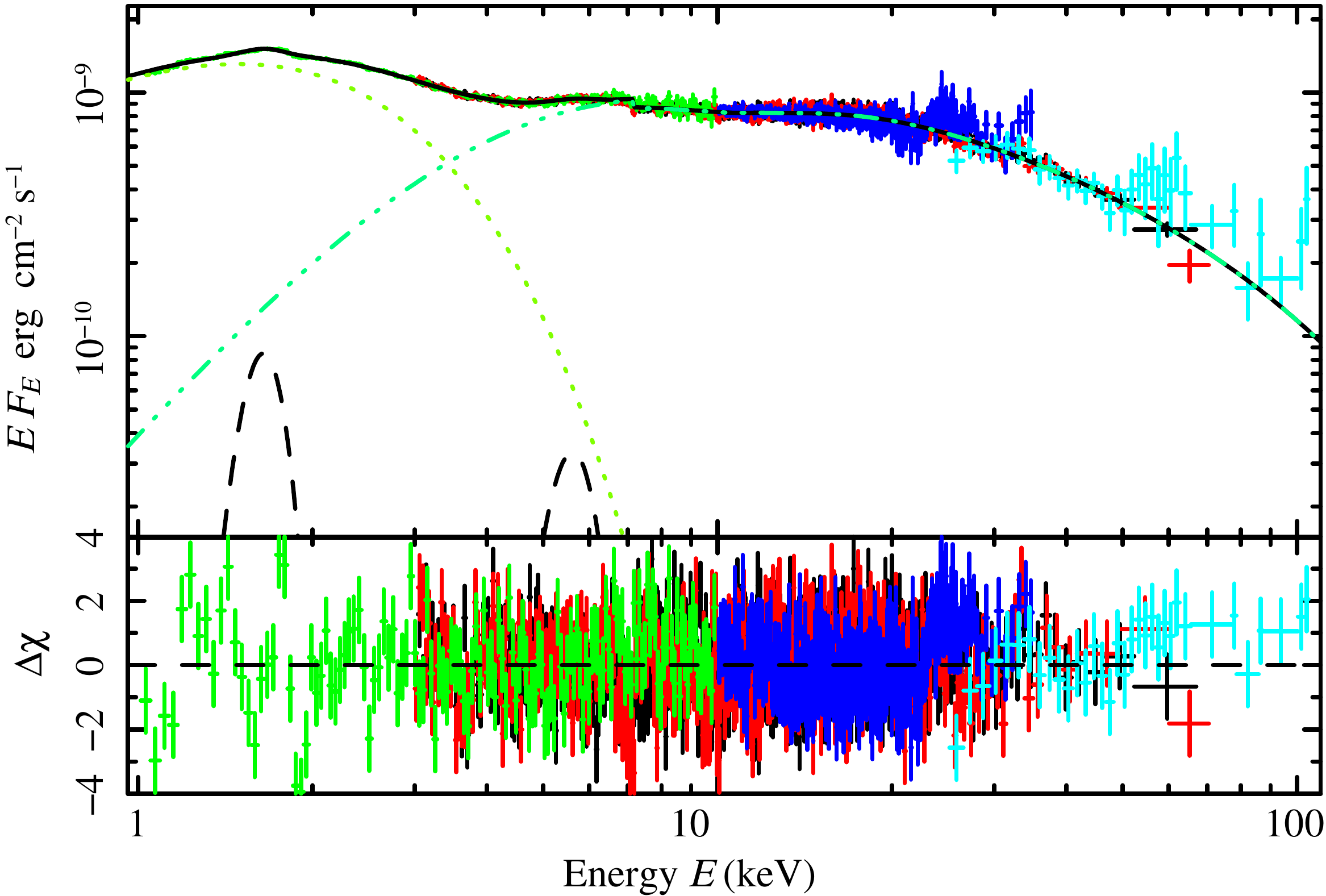}
\caption{The unabsorbed spectra  of \psrtar\ from \nicer, \nustar\ FPMA, \nustar\ FPMB, \hxmt\ ME and \hxmt\ HE in the 1--150 keV energy range. The fitting model is {\tt constant$\times$tbabs (gaussian+gaussian+diskbb+compps)}. The black dashed lines represent the two gaussian lines. The best model is plotted as black solid line. The disk blackbody and {\tt compps} components are shown in dotted and dash-dot-dot-dot lines, respectively. %The reduced $\chi^2$ is 1.17 for the total d.o.f of 2299. 
\label{fig:spec}}
\end{figure}

We noted that the only  \nustar\ observation was carried out quasi-simultaneously with some of \nicer\ and \hxmt\ observations. We fitted the  spectra, including the \hxmt\ ME and HE  data between MJD 	59753.46--59754.40 (obs. ID P0404275015) with a total exposure of 20.99 ks (ME) and 17.51 ks (HE),  \nustar\ FPMA (29.55 ks), FPMB (29.70 ks) and \nicer\ data starting on MJD 59753.30 (5.21 ks, obs. ID 5533011501). We kept the default energy range of 3--79 keV for \nustar\ FPMA/FPMB, and for the other instruments the same ranges as in Sect.~\ref{sec:nicer_spec}. The multiplication factors were also included to account for the cross-calibration uncertainties between the instruments.  The factor was fixed at unity for the \nustar\ FPMA, and set free for the other instruments. 
We first fitted the spectra by using an absorbed cutoff power-law model in combination with a disk black body model and two additional Gaussians. The best fitted model yielded $N_{\rm H}\approx2.30\times10^{22}$~cm$^{-2}$, a photon index $\Gamma\approx 1.54$, a cutoff energy $E_{\rm cut}\approx24.1$ keV, the \textbf{Gaussian} at 5.44 keV with a width of 1.32 keV, a disk temperature $kT_{\rm diskbb}\approx 0.61$ keV, and a reduced $\chi^2\sim1.15$ for a total d.o.f. of 2300. We then replaced the phenomenological cutoff power-law model by the thermal Comptonization model {\tt compps}, which is  {\tt tbabs (gaussian+gaussian+diskbb+compps)}. We obtained a best-fit reduced $\chi^2$ of 1.17. We noticed that a reflection feature has been reported by \citet{ATel15470}, therefore the reflection parameter, $R$, of {\tt compps} was set free.  The best-fit reduced \chiq{} decreased to 1.11 for a total d.o.f. of 2298, better than the cut-off power-law model.   The best-fit parameters are listed in Table~\ref{table:spec} and the broad-band spectrum is shown in Figure~\ref{fig:spec}. The \nicer\ spectrum contributed the most significant residuals.  
Considering such a large data set, this model provides an acceptable fit to the spectra. We obtained a normalization factor for the seed photons, $K_{\rm bb}=65.9\pm6.8$, which implies a size of blackbody emitting region $R_{\rm bb}=0.34\sqrt{K_{\rm bb}}\approx3 ~{\rm km}$ for a distance of 3.4 kpc  \citep[Wang et al. submitted, see also][]{Chen2022}. This region is smaller than the whole NS surface with a typical radius of 12 km. However, the true emitting size should be larger since part of photons were reflected by the accretion disk. It is worthy to note that it is difficult  to constrain the refection contribution by only using the joint \nicer\ and \hxmt/ME/HE spectra, i.e., without \nustar\ data, the $\chi^2_r$ is 1.0, see Sect.~\ref{sec:nicer_spec}.
% 

%In addition, the inclination angle was also thaw. 

We also changed the {\tt compps} model to a relativistic reflection model, {\tt relxillCp} \citep{Dauser16}, in which the reflection spectrum is calculated
using the Comptonization continuum  {\tt nthcomp}. After many attempts, during the fit, we set the iron abundance, $A_{\rm Fe}$, the density of the accretion disk, $\log N$,  the
ionization parameter, $\log \xi$, the power-law index of the incident spectrum, $\Gamma$, the reflection fraction, $R_f$, the outer emissivity index, $q_2$, free to vary. The spin
parameter was fixed at 0.248 for the NS spin frequency of 528 Hz. The outer disk radius was fixed at 1000\,$R_{\rm ISCO}$. The inclination angle, $i$, was also fixed at 60 degree. Other parameters were fixed at their default values. This model gave a best-fit reduced $\chi^2$ of 1.11 for a total d.o.f. of 2295, slightly larger than the {\tt compps} model. The results are listed in Table~\ref{table:spec}.

 \begin{table} %[h] 
 \caption{\label{table:spec} The spectral fitting to the broadband spectra of \psrtar\  with the \nicer/\nustar/\hxmt\, observations. The best-fitted results in the middle and right columns are from the  model  {\tt constant$\times$tbabs} {\tt (gaussian+gaussian+diskbb+compps)} and {\tt constant$\times$ tbabs} {\tt (gaussian+gaussian+diskbb+relxillCp)}, respectively.}
 \centering
 \begin{tabular}{lll} 
  \hline 
 \hline 

Parameter (units) & \multicolumn{2}{c}{Best-fit values}  \\

 \hline 
 $N_{\rm H}~(10^{22} {\rm cm}^{-2})$ & $ 2.25\pm0.01$ & $2.41\pm0.01$\\ 
 $E_{\rm line,~1}$   (keV)  & $1.62\pm0.02$ & $1.66\pm0.02$ \\ 
 $\sigma_1$   (keV)  & $0.13\pm0.01$ & $0.09\pm0.01$ \\ 
 Norm$_1$($\times10^{-3}$)     & $6.5\pm0.6$ & $4.2\pm0.7$\\ 
 $E_{\rm line,~2}$   (keV)  & $5.53\pm0.06$ &$5.78\pm0.05$\\ 
 $\sigma_2$   (keV)  & $0.48\pm0.07$ & $0.76\pm0.06$ \\  
 Norm$_2$$(\times10^{-4})$     & $8\pm2$ & $26\pm4$ \\ 
 Inclination (deg) & 60 (fixed)&60 (fixed) \\
 % $kT_{\rm bb,~refl.}$   (keV)  & $1.15\pm0.03$ \\ 
 $kT_{\rm diskbb}$   (keV)  & $0.63\pm0.01$ &$0.55\pm0.01$\\ 
 $K_{\rm diskbb}$   (km$^2$) & $770\pm38$ & $826\pm58$\\  
  $kT_{\rm e}$      (keV)   & $35.6\pm1.1$   & $16.8\pm0.6$\\  
  $q_2$ & - & $1.87\pm0.49$ \\
  $\Gamma$ & -& $1.88\pm0.01$ \\
  $\log \xi$ & -& $4.31\pm0.1$\\
  $\log N$ & -& $17.98\pm0.38$\\
 $kT_{\rm bb, ~seed}$   (keV)  & $1.31\pm0.03$ & - \\  
 $\tau_{\rm T}$     & $0.89\pm0.04$  & - \\ 
 $R$ &$1.34\pm0.09$ & $6.51\pm4.67$ \\ 
Norm$_{\rm refl.}~(\times10^{-4})$     & - & $8.3\pm1.2$ \\ 
 $K_{\rm compps}$    (km$^2$)  & $65.0\pm6.8$     & - \\ 
 $C_{\rm \nustar/FPMA}$ & 1 (fixed) &1 (fixed)\\ 
 $C_{\rm \nustar/FPMB}$ & $1.01\pm0.01$ &$1.01\pm0.01$ \\ 
 $C_{\rm \nicer}$ & $1.05\pm0.01$ & $1.05\pm0.01$\\ 
 $C_{\rm \hxmt/ME}$ & $1.11\pm0.01$ & $1.11\pm0.01$\\ 
 $C_{\rm \hxmt/HE}$ & $1.20\pm0.03$ & $1.21\pm0.05$\\ 
  \hline 
 $\chi^{2}/{\rm d.o.f.}$ &  2543.6/2298 & 2550.9/2295\\ 
 $F_{\rm bol}$ ($10^{-9}$ erg s$^{-1}$ cm$^{-2}$)\tablenotemark{a} & $3.97\pm$0.01 & $3.96\pm$0.01  \\ 
 \hline  
 \end{tabular}  
 \tablecomments{
Best parameters determined from the fits to the average
 broad-band spectrum of \psrtar\  performed with the \nicer/\nustar/\hxmt\ data. The multiplication factor for all instruments are provided.
 \tablenotetext{a}{ Unabsorbed flux in the 1--250 keV energy range.}}
% \tablefoottext{a}{Unabsorbed flux in the 0.1--250 keV energy range.}}
 \label{tab:table3} 
 \end{table}

\section{Summary and discussion}\label{sec:discussion}

In this paper we reported on detailed timing and spectral analyses of \hxmt,  \nicer\ and \nustar\ data collected from the newly confirmed AMXP \psrtar\ during its 2022 outburst.  The outburst of \psrtar\ reached its peak flux within the first few days at $\sim 1\times10^{-8}~{\rm erg~cm^{-2}~s^{-1}}$ and decayed into quiescent state in the next $\sim 25$ days. The outburst profile observed from \psrtar\ is quite similar to that of other AMXPs, showing a fast rise ($\sim$1--2 days), an exponential followed by a linear decay, which can be explained by the disk instability model \citep{king1998, Powell2007}.

During the full outburst hard X-ray/soft $\gamma$-ray emission was detected up to $\sim 200$ keV. The joint \nicer\ and \hxmt\ spectra in the energy of 1--150 keV were well fitted by a combination of a multi-color disk black body and Comptonization model {\tt compps}, with two additional Gaussian lines, exposed to interstellar extinction. A joint spectral fit including quasi-simultaneous \nustar, \nicer\ and \hxmt\ ME/HE data showed  emission reflected from the accretion disk, which can be modelled by the reflected {\tt compps} model or the {\tt relxillCP} model. 

%a blurred (0.48 keV) discrete spectral feature at 5.53 keV compatible with the Iron K$\alpha$ line modified by the relativistic Doppler effect due to the fast motion of the matter in the inner accretion disk \citep[see also][]{DiSalvo19,Papitto13d}. The spectra also showed

Usually, the hard X-ray emission from  AMXPs is relatively weak and has typically pulsation amplitudes of less than 10\% 
(see e.g., \citealt{papitto20}, and references therein; see also \citealt{falanga05} for IGR J00291+5934 with the pulsed fraction of $\sim12-20\%$ at $\sim100$ keV). Therefore, the pulsations of most AMXPs are mostly detected at soft X-rays with instruments aboard e.g \rxte, XMM-Newton or \nicer. 
At hard X-rays, \hxmt\ observed Swift J1756.9--2508 during its 2018 outburst, and detected the X-ray pulsations up to $\sim100$ keV using an ephemeris determined by \nicer\ at soft X-rays \citep{ZLi21}. 
In the case of \psrtar, at hard X-rays both the ME and HE aboard of \hxmt\ were capable of deriving the pulsar ephemeris directly from their event data, obtaining parameters well consistent with those derived by \nicer\ at soft X-rays.
Particularly, our timing analysis using the DE405 solar system ephemeris and the most accurate VLA \citep[][]{ATel15481} source location for \psrtar\ in the barycentering process, contrary to \citet{Bult22} who used DE430 and the less accurate ($\sim 0\farcs4$ difference) \swift\ location \citep[][]{ATel15421}, yielded consistent orbital and spin parameters with \citet{Bult22} at $1\sigma$-confidence.

From \hxmt\ observations hard X-ray pulsations have been detected significantly up to $\sim 95$ keV. The pulse profiles in 0.8--95 keV band are well described by a truncated Fourier series. The phase angles of the fundamental and the first overtone are nearly constant for energies below $\sim 60$ keV, implying that most photons are likely emitted from the same region at the NS surface with a very similar emission pattern. The observed pulse profiles could be used to measure the NS mass and radius \citep{Poutanen03} of \psrtar\ in a future study. 
% However, the harder photons reach the observer later than the soft ones, which could be related to the fact that these photons undergone many more scatterings and took longer time to arrive to the observer \citep{Falanga07}. 

The stability of the pulse profiles across a broad energy range allows us to determine the absolute timing accuracy of \hxmt\ in detail. Cross-correlating \hxmt\ ME and \nicer\ pulse profiles for the overlapping 5--10 keV energy band yielded a phase shift $\delta\Phi$ of only $-0.008\pm 0.005$, consistent with zero, suggesting that both instruments are well aligned within 14~$\mu$s, with the \hxmt\ ME events arriving a little bit later than \nicer\ ones. A similar correlation analysis now between \hxmt\ ME and HE pulse profiles for the 20--35 keV band resulted in a difference of only 0.9~$\mu$s with an uncertainty of $13.9~\mu$s.

\citet{Bult22} performed a coherent timing analysis adopting a constant frequency- and a flux bias model using \nicer\ ToA’s
covering the {\em full} outburst period. They found unacceptably high $\chi^2$ values for both models assumptions resulting in
complex timing residuals for the first 4–5 days during the rising part of the outburst, a flat shape during the next $\sim 20$ days in the decay
phase, and finally, (linearly) increasing phase deviations at the end of outburst, also visible at the very beginning of the outburst. 

In this work we analysed the timing data from \hxmt\ ME and HE observations as well as from \nicer, and found that 
the ToAs from the different (independent) instruments were well consistent with each other in spite their 
different bandpass. The timing residuals showed erratic behavior (or V-shaped wedge) during the rising part of the 
outburst between MJD 59737.0 and 59741.9, similar to the results shown in \citet{Bult22}.

For the decaying part of the outburst between MJD 59741.9--59760.6 we detected a significant positive quadratic term 
$\dot{\nu}$ of $(9.0\pm2.1)\times10^{-14}~\rm{Hz~s^{-1}}$ fitting a two-parameter 
timing model, including $\nu$ and $\dot{\nu}$, yielding an acceptable $\chi^2$, which could interpreted as a spin-up.

We realize that such spin-up or spin-down terms can arise when an inaccurate source location is
used in the barycentering process. However, in our work we have used the sub-arcsecond accurate VLA-location for \psrtar, with an accuracy better than $0\farcs005$. This uncertainty on the location can account for an uncertainty in the 
spin-up rate \citep[see e.g.,][for the formulae]{Sanna20} of only $\sim2.0\times10^{-16}~{\rm Hz~s^{-1}}$, 2.7 order 
of magnitude smaller than the measured $\dot{\nu}$ value, and much smaller than our error estimate of 
$2.1\times10^{-14}~\rm{Hz~s^{-1}}$.
On the contrary, if we should have used the \swift-XRT location in the barycentering process with a centroid $0\farcs4$ 
from the VLA-location and an estimated error radius of $2\farcs2$ then the uncertainty in the quadratic term would be 
$\sim$80 times larger, i.e. of the order of $1.6\times10^{-14}~\rm{Hz~s^{-1}}$,  comparable to our measurement error.
Therefore, we assume that the measured value for $\dot{\nu}$ during the decay phase is accurate and genuine, and 
could be interpreted as a gain of angular momentum by the NS from the accreted matter.

During the outburst X-ray pulsations have been detected at bolometric flux states with a flux between $\sim(0.1-1)\times10^{-8}~{\rm erg~cm^{-2}~s^{-1}}$. Adopting a source distance of 3.4 kpc, as constrained by Wang et al. (2023, submitted, see also \citealt{Chen2022}), these bolometric fluxes correspond to accretion rates in the range  $(0.012-0.12)\dot{M}_{\rm Edd}$, where $\dot{M}_{\rm Edd}=2\times10^{-8}~M_\odot~{\rm yr^{-1}}$ is the Eddington rate, assuming that $ L=\eta \dot{M}c^2$ and a typical value of $\eta=0.1$ for NS \citep{frank02}.
By adopting Equations 11 and 12 and the same assumptions in \citet{Psaltis99b}, these limits allow us to constrain the surface dipole magnetic field strength of $(0.25-74)\times10^8$~G \citep[see also][]{Hartman2008}. If 20\% uncertainty of distance is considered, the  magnetic field is in the range of   $(0.2-85)\times10^8$~G. The possible bolometric correction factor of the flux would not change the magnetic field significantly because the measured fluxes are covered a broadband energy range.

During the spin-up period the averaged bolometric flux is $0.56\times10^{-8}~{\rm erg~cm^{-2}~s^{-1}}$, corresponding to $0.065 \dot{M}_{\rm Edd}$. If the spin-up of the pulsar is solely induced by the angular momentum transferred from the accreted material to the NS, the magnetic field can be determined from the relation \citep{Shapiro83, Tong15, Pan22}, 
\begin{equation}
B=0.1I^{7/2}_{45}R_6^{-3}M_{1.4}^{-3/2}\dot{\nu}_{-13}^{7/2}\left( \frac{\dot{M}}{0.065\dot{M}_{\rm Edd}}\right) ^{-3}\times10^8~{\rm G},
\end{equation}
where $I_{45}$, $R_6$, $M_{1.4}$,  $\dot{\nu}_{-13}$ are the moment of inertia, the NS radius and mass, the spin up in units of $10^{45}$~g~cm$^2$, $10^6$~cm, $1.4M_\odot$, and $10^{-13}$~Hz~s$^{-1}$, respectively. By adopting $I_{45}=1.5$ \citep[see e.g.,][]{Worley08} and the measured spin-up rate, we obtain a magnetic field strength of $(0.04-2)\times10^8$~G. Combined with the aforementioned range, the magnetic field strength of \psrtar\ is $(0.2-2)\times10^8$~G,  which is comparable with magnetic field strengths estimated for several other AMXPs \citep[see e.g.,][]{Hartman2008,Patruno09c,Patruno2010,Hartman2009,Hartman11,Papitto11}. 

%%%%%%%%%%%%%%%%%%%%%%%%%%%%%%%%%%%%%%%%%%%%%%%%%%%%%%%%%%%%%%%%

%\begin{acknowledgments} 
\section*{}

We appreciate the referee for all valuable comments and suggestions that led to the improvement of the manuscript.
This work was supported the Major Science and Technology Program of Xinjiang Uygur Autonomous Region (No. 2022A03013-3). Z.S.L., Y.Y.P. and L.J. were supported by National Natural Science Foundation of China (12103042, U1938107, 12273030, 12173103). S.Z. is supported by the National Key R\&D Program of China (2021YFA0718500).
J.P. acknowledges support from the Academy of Finland grant 333112.
This work is also supported by International Partnership Program of Chinese Academy of Sciences (Grant No.113111KYSB20190020). This work made use of data from the \hxmt\
mission, a project funded by China National Space Administration (CNSA) and the Chinese Academy of Sciences (CAS), and also from the High Energy Astrophysics Science Archive Research Center (HEASARC), provided by NASA’s Goddard Space Flight Center. 
%\end{acknowledgments}

%%%%%%%%%%%%%%%%%%%%%%%%%%%%%%%%%%%%%%%%%%%%%%%%%%%%%%%%%%%%%%%%


\vspace{5mm}
 \facilities{\hxmt, \nicer, \nustar}
% \software{}


\bibliography{pulsars}{}
\bibliographystyle{aasjournal}

\end{document}